# Exploring Iron-Arsenide monolayers in 1T and 1H phases as anode materials for Lithium-ion batteries: A first-principles study

Ajay Kumar[a] and Prakash Parida[a]*
[a]Department of Physics, Indian Institute of Technology Patna, Bihta, Bihar, India
*Corresponding authors: - [a]pparida@iitp.ac.in

**Abstract**

This theoretical investigation delves into the structural, electronic, and electrochemical properties of two hexagonal iron-arsenide monolayers, 1T-FeAs and 1H-FeAs, focusing on their potential as anode materials for Lithium-ion batteries. Previous studies have highlighted the ferromagnetic nature of 1T-FeAs at room temperature. Our calculations reveal that both phases exhibit metallic behaviour with spin-polarized electronic band structures. Electrochemical studies show that the 1T-FeAs monolayer has better ionic conductivity for Li ions than the 1H-FeAs phase, attributed to a lower activation barrier of 0.38 eV. This characteristic suggests a faster charge/discharge rate. Both FeAs phases exhibit comparable theoretical capacities (372 mAhg$^{-1}$), outperforming commercial graphite anodes. The average open-circuit voltage for maximum Li atom adsorption is 0.61 V for 1H-FeAs and 0.44 V for 1T-FeAs. The volume expansion over the maximum adsorption of Li atoms on both phases is also remarkably less than the commercially used anode material such as graphite. Further, the adsorption of Li atoms onto 1H-FeAs induces a remarkable transition from ferromagnetism to anti-ferromagnetism, with minimal impact on the electronic band structure. In contrast, the original state of 1T-FeAs remains unaffected by Li adsorption. To summarize, the potential of both 1T-FeAs and 1H-FeAs monolayers as promising anode materials for Lithium-ion batteries, offering valuable insights into their electrochemical performance and phase transition behaviour upon Li adsorption.

## 1. Introduction

Rechargeable lithium-ion batteries (LIBs) play a crucial role in various energy devices, ranging from portable electronics to driving the growth of electric vehicles.[1-3]. Rechargeable LIBs are essential for energy storage, providing adaptable solutions for emerging industries. These batteries have demonstrated their reliability as power sources for notebook computers, mobile phones, and digital products, contributing significantly to technological advancements.[4, 5]. However, limited energy density has hindered the rapid development of hybrid and electric vehicles [6-8]. Addressing this constraint requires urgently exploring new materials with high capacity for anode electrodes, particularly those featuring a better voltage platform and superior energy density.

The novelty of two-dimensional materials is due to the high surface-to-volume ratio and the restriction of electronic motion in one direction, resulting in quantum effects[9-11]. Due to these two major factors, monolayers exhibit distinct and innovative features compared to the bulk, attracting researcher's attention and encouraging them to investigate other low-dimensional materials. Two-dimensional materials have several uses, including energy storage and conversion, health and environmental monitoring, and medicines[12-18]. Their structure and morphology are easily customizable, increasing their understanding of basic physical processes and application performance. Since graphene[19] was discovered in 2004, various 2D materials, such as boron nitride[20], silicene[21], borophene[22, 23], stanene[24], transition metal dichalcogenides (TMDs)[25, 26], MXenes[27, 28], and others, have been fabricated in experiments. TMDs often have many atomic layers, allowing them to stay magnetically stable, whereas single atomic layers of metal-based materials are rare. Further, the novelty of TMDs is that the number of electrons can provide a route for long-distance magnetic interaction. Generally, heavier transition metals in TMD provide higher anisotropy, essential to maintaining the magnetic order in two-dimensional materials. Iron-based compounds as a superconductors are well-studied theoretically as well as experimentally[29]. First, the superconductive nature of iron compound was $Th_7Fe_3$, reported in 1961 by Matthias, Compton, and Corenzwit with a critical temperature ($T_c$)=1.8K. After that, A lot of structures were reported, like LaFePO[30], LaFeAsO[31], MFeAs family[32], and iron chalcogenide FeSe family[25], and other doped Iron compounds were reported to improve the $T_c$ [33-36]. The single-layer FeAs trigonal phase is identical to the LaFeAsO structure, first reported by Jun Dai in 2009 [37]. It is reported that single-layer FeAs cannot maintain the collinear antiferromagnetic state along the z-axis, yet ferromagnetic states are still conceivable. Additionally, they tried to find the FeAs layer in ReFeAsO and $AFe_2As_2$ by ignoring the interaction between the ReO (A) and FeAs single layer but failed to find the correct geometry. In 2019, the three different phases of ferromagnetic FeAs monolayers at room temperature were recently reported by Yalong Jiao[38]. He reported one of the three is the trigonal phase (rhombus unit cell) along with two Tetragonal lattices where the former is less stable than the latter two geometry. In the trigonal phase, one Fe atom seems to occupy the centre of one octahedron. In contrast, a second Fe atom shares one of its corners and will eventually take the centre of another octahedron. Therefore, the reported trigonal structure is 1T-FeAs phase. 1T-FeAs is MX type of structure which is well studied for group-III monochalcogenides solar energy conversion and water splitting applications[39-41]. Is there any other phase of the FeAs combination yet to explore because metal compounds exist in multiple phases depending on the temperature and other parameters? For illustration, $MoS_2$ and other TMDs show $1T-MX_2$ and $1H-MX_2$ or $2R-MX_2$ phases of monolayers has been reported[42, 43]. Similar approach has been used to investigate the comparative study of the 1H- and 1T-MX phases of FeAs monolayer. Further, Iron-based monolayers shows impressive electrochemical properties as an electrode material for Li and non-Li ion batteries[44-46]. Therefore, it is also interesting to explore the electrode like features of iron-arsenide phases.

In this work, a systematic study has explored the structural, electronic, magnetic and electrochemical properties of the 1T-FeAs and 1H-FeAs phases. A comparative study to check the stability of FeAs structures has been done regarding their formation energies and phonon dispersion relations. Further, electronic and magnetic properties have been revealed with spin-polarization calculation using GGA+U functional. Spin polarization FeAs monolayers may be crucial in spintronics, nanoelectronics, renewable energy, and other high-tech fields. Moreover, the electrochemical characteristics of FeAs monolayers as anode materials have been investigated for Li-ion batteries. The electrochemical parameters like adsorption energy, charge density difference, diffusion barrier, theoretical specific capacity and open circuit voltage have been evaluated for anode material characteristics of FeAs phases.

## 2. Computational details

Vienna Ab Initio Simulation Package (VASP) is used to study the FeAs and Li-FeAs structure using Density Functional Theory (DFT). It employs the Projector Augmented Wave (PAW) approach, emphasizing interactions between valence and core electrons and periodic boundary conditions. The pseudopotentials of Li, Fe and As have electronic configurations with valence electrons are $1s^1 2s^1$, $3d^6 4s^2$ and $4s^2 4p^3$, respectively. For the exchange-correlation potential defined by Perdew–Burke–Ernzerhof (PBE), we use Generalized-Gradient approximation (GGA+U). The plane wave energy cut-off value and K-mesh grid values for optimization and self-consistent calculations are 600 eV and 11×11×1, respectively. The conjugate-gradient approach fully relaxes the suggested crystal structures with a force value of $10^{-3}$ eV/Å per atom. Further, the energy convergence criteria are $10^{-8}$ eV throughout the calculations. Both structures have rhombus unit cell that follows a periodic pattern in the x-y plane, and a vacuum of 20 Å is applied to prevent interaction along the Z-axis. Both the structures have the same high symmetry points, Γ−M−K–Γ, to calculate electronic band spectra. The dynamical stability of these systems was identified using phonon dispersion calculations. We use the finite displacement approach to execute the vibrational spectra of atoms by phonopy package. To calculate phonon dispersion spectra, we build supercell 3×3×1 for both structures and calculate force constants with the conjunction between VASP and phonopy. In order to study the electrochemical properties, the van der Waals (vdW) interactions between the adatoms with monolayer have been taken into consideration by using the Grimme zero damping DFT-D3 method. Further, to determine the transition state (TS) structure and diffusion barrier of Li-adsorbed FeAs monolayers, the climbing image nudged elastic band (CI-NEB) method has been used. We utilize VESTA (Visualization for Electronic Structural Analysis) to prepare and visualize the crystal structure. Xcrysden is used to display the Brillouin zone, which helps to find the high symmetry path for band structure plots.

## 3. Results and discussions
### 3.1. Crystal structure, structural stability and phonon dispersion calculation

We recalculate the lattice parameters of the 1T-FeAs lattice, which had previously been studied theoretically by Yalong Jiao, 2019. The top and side views of the different phases of FeAs are depicted in **Figure 1**, where (a) represents the trigonal prismatic coordination (H) of FeAs and (b) possesses the octahedral (T) coordination pattern of FeAs. DFT simulations confirmed the trigonal *p3m1* lattice symmetry of a relaxed primitive cell of 1T-FeAs with four atoms, which is in good agreement with the previous report [30].

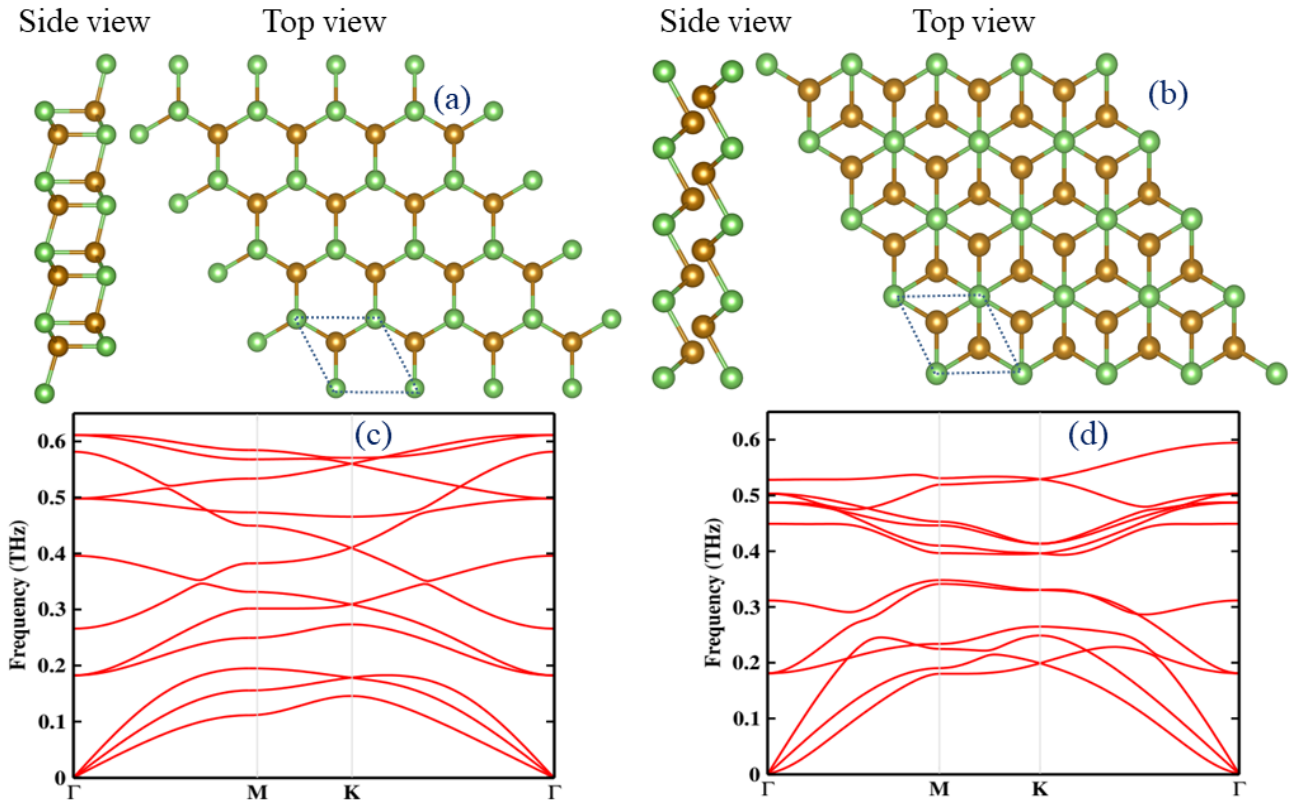

**Figure 1**: *Atomic structures of FeAs monolayers (a) 1H and (b) 1T phases, with a primitive unit cell indicated by a dot rhombus. Fe and As atoms are colour-coded by Brown and Green, respectively. The phonon dispersion plots of 1H-FeAs and 1T-FeAs are shown in (c) and (d).*

On the other hand, the lattice parameters of 1H-FeAs as a=b=3.65 Å. The 1H-FeAs phase is similar to honeycomb boron nitride (h-BN), with alternate Fe-metal and As-atoms. The main difference is that in h-BN, all atoms sit in a single plane, but four separate atomic layers lie along the z-axis here. Like the 1T-FeAs, 1H-FeAs have a trigonal $p3m1$ lattice symmetry and four atoms in each primitive unit cell. Both the phases have four atomic layers, As-Fe-Fe-As with slightly different thicknesses (1T-FeAs has 2.73 Å and 1H-FeAs has 3.39 Å) along the z-direction, whose top and side views are shown in **Figure 1(a)** & **1(b).**

To examine the thermodynamics stability for both phases, we have found the cohesive energy ($E_C$) for 1T-FeAs and 1H-FeAs by using the following relation,

$$E_C = E_{FeAs} - (2E_{Fe} - 2E_{As})/4 \qquad (1)$$

Where $E_{FeAs}$, $E_{As}$ and $E_{Fe}$ are the energy of single-layer FeAs, As atom and Fe atom, respectively. The cohesive energy for 1T-FeAs is -6.14 eV/atom, and for 1H-FeAs is -3.42 eV/atom. These negative values of $E_C$ indicates that the synthesis of FeAs monolayers is energetically favoured.

Further, for dynamic stability of both the FeAs monolayers has been verified by calculating the phonon dispersion band spectra. For this, the finite displacement method has used a supercell of 3×3×1 for both FeAs phases to calculate the phonon dispersion calculations. **Figures 1(c)** and **1(d)** reveal that the 1T-FeAs and 1H-FeAs are dynamically stable and can exist as a freestanding 2D crystal because the imaginary frequencies are absent in the Brillouin zone. Further, the maximal phonon frequency of 1T-FeAs (5.96 THz) and 1H-FeAs (6.15 THz) is due to the significantly heavier nuclei of As and Fe atoms. In **Figure 1(c)**, it has been observed that a gap of 0.531 THz between the optical branches of 1T-FeAs in the middle-frequency region, whereas 1H-FeAs has highly dispersed spectra which include a Dirac cone-like linear crossing of phonon bands. The more dispersive phonon spectra of 1H-FeAs show that it has more phonon-phonon scattering because of strong acoustic and optical band coupling[47].

### 3.2. Electronic properties

**Figure 2** depicts the spin-polarized electronic band structures of 1H-FeAs and 1T-FeAs. Both FeAs monolayers are metallic because of up-spin and down-spin bands crossing the Fermi energy.

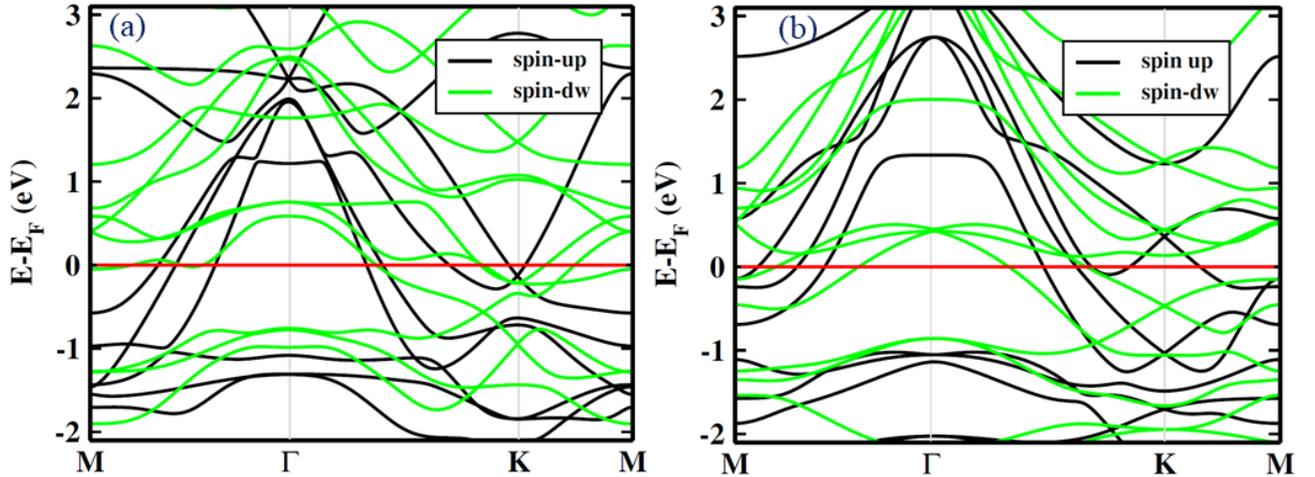

**Figure 2:** *Spin-polarised electronic band spectra of (a) 1H-FeAs and (b) 1T-FeAs.*

In **Figure 2(a)**, the spin-up bands (black colour) of 1H-FeAs look to meet at Γ point at around 2.3 eV above the fermi energy, whereas the spin-down (green colour) bands are dispersed uniformly. Notably, near Γ the point, a downward concave pattern becomes flat around 1.24 eV in both the FeAs monolayers. A similar upward concave curve can be seen around the M point. Further, there are several topological band crossings at the K-point of the Brillouin zone in various energy levels. In the 1T-FeAs phase, spin-up bands cross each other near the fermi energy at K-points. Similarly, in 1T-FeAs, spin-up bands show linear crossing at 0.34 eV above and 1.03 eV below the Fermi energy, whereas linear response can be found at 0.48 eV below and 1.26 eV above Fermi energy for a spin down. Comparatively, 1H-FeAs also show an identical pattern of downward concave near Γ point and upward concave dispersion of bands near M high symmetry points.

The spin-polarised projected density of states (DOS) for both the FeAs phases are shown in **Figure S1** in the supporting information (SI). Notably, the spin-up bands in proximity to the Fermi energy primarily arise from the *p*-orbital of As, accompanied by a minor contribution from $d_{xz}$ and $d_{x^2-y^2}$ orbitals of Fe. Conversely, the spin-down bands exhibit the main involvement of $d_{x^2-y^2}$ and $d_{z^2}$. Likewise, the DOS projection for 1T-FeAs (depicted in **Figure S1 (b)**) is characterized by a significant contribution from the d-orbital of Fe and the p-orbitals of As. Specifically, the spin bands are chiefly influenced by the *p*-orbitals of As, with minor contributions from $d_{xz}$ and $d_{x^2-y^2}$. In contrast, the spin-down bands are primarily shaped by $d_{z^2}$, $d_{x^2-y^2}$, and $d_{xz}$ orbitals of Fe with a minor contribution of the *p*-orbitals of As. The inherent metallic nature of both FeAs monolayers confers a distinct advantage in terms of electrical conductivity and favourable electrochemical properties, ultimately enhancing the overall performance of battery cycling.

### 3.3. Li-FeAs as anode material
#### 3.3.1. Absorption of Li atoms on 1H-FeAs and 1T-FeAs

To assess the diffusion and storage capacity of Li atoms on the surface of both the FeAs monolayer, we focus on identifying the favoured adsorption sites for Li atoms in 1H-FeAs and 1T-FeAs phases. The adsorption energy has been calculated for a single Li atom adsorbed at different possible sites of the FeAs monolayer. We are looking for the most stable Li adsorption site on both FeAs monolayers.

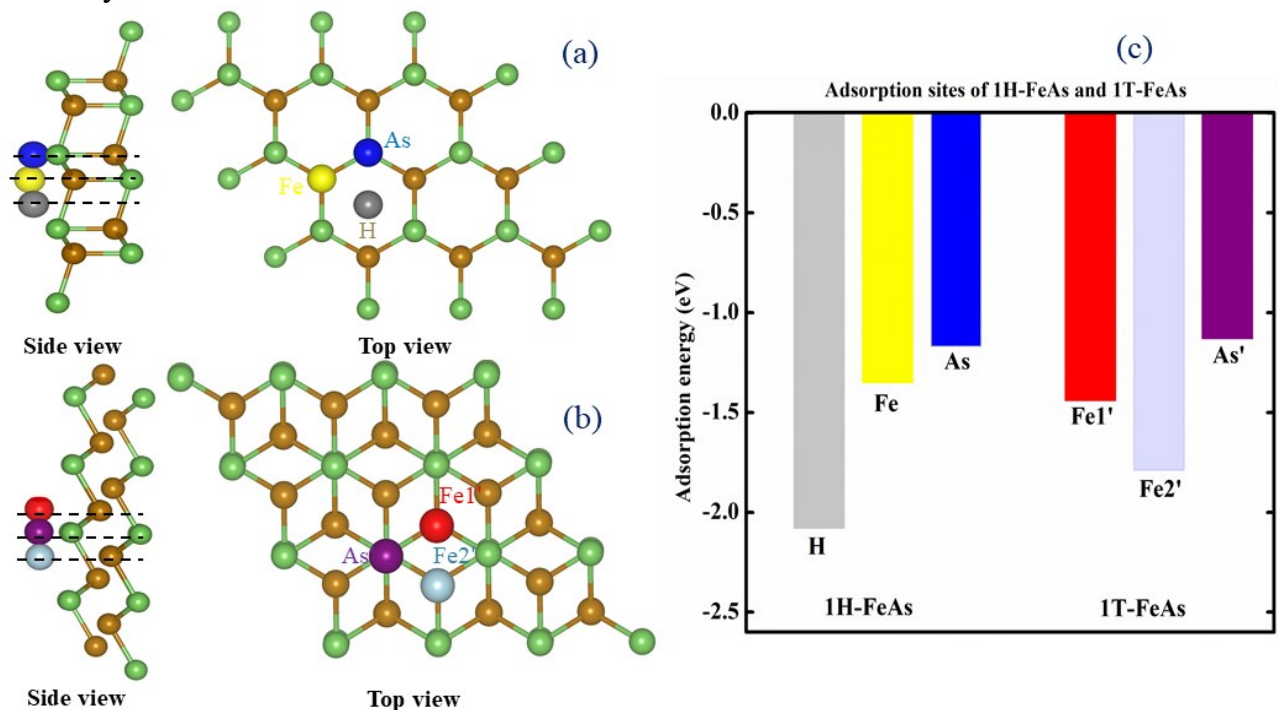

**Figure 3:** *(a) and (b) depict the possible adsorption sites for 1H-FeAs and 1T-FeAs monolayers, and 3(c) shows their corresponding adsorption energy at different sites.*

The 1H phase exhibits three distinct high-symmetry positions: the top position of the Fe atom corresponds to the Fe-site, the top position of the As atom corresponds to the As-site, and the apex of the hexagon at its centre represents the H-site, all highlighted in **Figure 3(a)**. Similarly, the 1T-FeAs phase features two distinct sites for Fe atoms: one for the upper layer, denoted as Fe1'-site, and one for the lower layer, denoted as Fe2'. Another site is located at the top of the As atom, referred to as the As'-site, as illustrated in **Figure 3(b)**. It is important to note that the prime symbol (')has been used to distinguish the 1T-FeAs sites from those in 1H-FeAs. The adsorption energy ($E_{ads}$) has been calculated as the difference between Li atom-adsorbed FeAs and the pristine FeAs and single atoms for both phases. Mathematically,

$$E_{ads} = E_{Li@FeAs} - E_{FeAs} - E_{Li} \qquad (2)$$

$E_{Li@FeAs}$ is the energy of Li-adsorbed FeAs monolayer, $E_{FeAs}$ is the energy of pristine FeAs monolayer and $E_{Li}$ is the energy of single Li atom.

In order to neglect the Li atoms interaction, a supercell of dimensions 3×3×1 has been used. In nearly all sites on the FeAs monolayer, effective Li atom adsorption is observed, indicated by the negative values shown in **Figure 3(c)**. For 1H-FeAs, the H-site emerges as the most energetically favoured adsorption site due to its significantly negative adsorption energy of -2.08 eV. The Fe-site exhibits an adsorption energy of -1.35 eV, and the As-site has -1.17 eV. Conversely, in 1T-FeAs, the adsorption energy for Li doping is -1.44 eV at the Fe1'-site, -1.79 eV at the Fe2'-site, and -1.12 eV at the As'-site. Notably, the adsorption energy at the top of the Fe atom is approximately 0.6 eV lower than that at the top of the As atom. In conclusion, the top of the hollow site in 1H-FeAs and the top of the Fe atom in 1T-FeAs represent the most energetically favourable sites for Li atom adsorption.

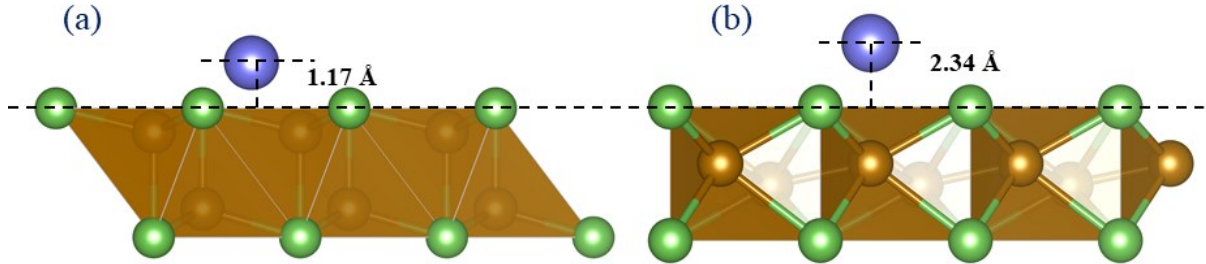

**Figure 4:** *The optimized height of Li atom from the surface of (a) 1H-FeAs and (b) 1T-FeAs monolayers.*

The optimized structures for the Li-adsorbed FeAs monolayers at the most preferable H-site in 1H-FeAs and C-site for 1T-FeAs are shown in **Figure 3(a)** and **(b),** respectively. The optimized interaction distance between the Li atom and the FeAs monolayer is expressed in terms of the adsorption height (h), which is the vertical distance between a FeAs layer and the Li atom. The adsorption height depends on the electrostatic interaction between the FeAs sheets and Li atoms. It has been observed that adsorption height is smaller for the centred position than at the top of atomic sites. The adsorption energy at various sites with optimized vertical heights for both FeAs phases is reported in Table 1. It has been emphasized that a small atomic radius leads to higher adsorption energy and provides a strong adatom attachment to the surface of FeAs.

We further investigated the electronic characteristics of FeAs monolayers following the adsorption of Li atoms under the most energetically favourable sites. This investigation is crucial because electrode materials need to retain their conductive properties even after the adsorption of Li atoms. Specifically, we focused on the energetic favourable Fe2'-site of 1T-FeAs and the H-site of 1H-FeAs. In **Figure S2**, the electronic band spectra of Li-adsorbed FeAs reveal that the metallic nature of both the FeAs monolayers remains preserved after the adsorption of Li atoms. The adsorption of Li atoms causes a negligible alteration to the electronic band structure. The projected density of states (DOS) for Li-FeAs phases, illustrated in **Figure S3**, highlights the contribution of Li near the Fermi energy. Particularly, the s-orbital exhibits a small contribution, facilitating the occupation of an additional band by introducing an extra electron through the upward shift of the Fermi energy. The well-preserved metallic nature emphasizes robust electrical conductivity, offering substantial advantages for the performance of Li-ion batteries.

### 3.3.2. Charge Density difference

The charge density difference plots illustrate the behaviour of Li atoms at the H-site for 1H-FeAs and at the Fe2'-site for 1T-FeAs, as shown in **Figures 5(a) and 5(b),** respectively. The charge density difference plots for Li adsorption at alternate sites (Fe and As) in 1H-FeAs and (Fe1' and As'-site) in 1T-FeAs are presented in **Figures S4** and **S5**. The following formula has calculated the charge density difference between pristine and Li-adsorbed FeAs monolayers.

$$\rho_{net} = \rho_{(Li@FeAs)} - (\rho_{FeAs} + \rho_{Li}) \quad (3)$$

where $\rho_{(Li@FeAs)}$ and $\rho_{FeAs}$ are the charge densities of the optimized Li-adsorbed and pristine FeAs monolayer, respectively, and $\rho_{Li}$ is the charge density of the Li atom. These plots reveal distinct regions of electron accumulation (indicated in yellow) between the Li atoms and the FeAs monolayer. In contrast, electron depletion regions (indicated in blue) are observed around the Li atoms. These observations suggest that Li atoms act as electron donors, while the FeAs layers serve as electron acceptors. In other words, the adsorbed Li atoms transfer a fraction of charge to the FeAs monolayer. For quantitatively, Bader charge analysis has been done, and it has been found that Li transfer 0.65$e^-$ charge to 1H-FeAs and 0.57$e^-$ for 1T-FeAs. **Figure 5** shows that charge transfer is not localized over a single site but diffused over the surface of the FeAs sheet. For the H-site, the Li atoms interact with three As atoms at a distance of 2.59 Å, whereas for 1T-FeAs, the Li feels slightly less interaction with the As atom. Similarly, the values of charge transfer for other sites of both the FeAs phases have been reported in **Table 1**.

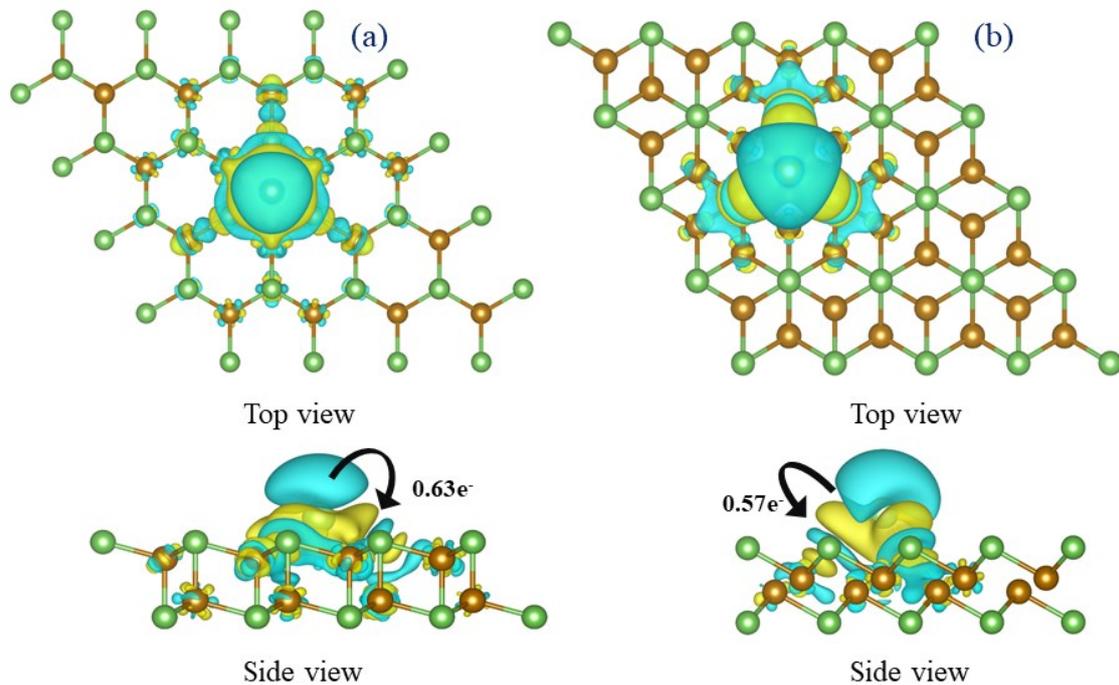

**Figure 5**: *The charge density plots of Li adsorbed at (a) H-site in 1H-FeAs and (b) C-site in 1T-FeAs with an iso-surface of 0.002e/Å³*

**Table 1** *reported the adsorption energy, height, and charge transfer at favourable possible sites for both the FeAs monolayers.*

| Phases | Sites | $E_{ads}$ (eV) | Height (Å) | Charge transfer (e) |
|---|---|---|---|---|
| 1H-FeAs | H-site | -2.08 | 1.17 | 0.67 |
| | Fe-site | -1.35 | 2.43 | 0.51 |
|  | As-site | -1.17 | 1.39 | 0.57 |
|  | Fe1'-site | -1.44 | 2.24 | 0.53 |
| 1T-FeAs | Fe2'-site | -1.79 | 1.37 | 0.44 |
|  | As'-site | -1.12 | 1.42 | 0.49 |

### 3.3.3. Diffusion of single Li atom on the 1T-FeAs and 1H-FeAs nanosheet

The rate of charging and discharging in this context largely depends on the electrical conductivity of both the FeAs nanosheets and the diffusion of Li ions. Therefore, we need to estimate the Li diffusion barriers to assess the diffusion of Li atoms on the surface of FeAs monolayers. In this regard, the Li diffusion barriers have been carried out on both the FeAs monolayers. Due to the inherent symmetry of the 1H-FeAs monolayer, the H-site has been found to be the most energetically favourable among the three adsorption sites for 1T-FeAs. Consequently, three different diffusion routes connecting two adjacent stable H-sites have been considered, as depicted in **Figure 6(a):** path-I, involving the migration of the Li atom across a Fe-As bond; path-II (H-Fe-H), involving the migration across over the Fe atom; and path-III (H-As-H), passing over the As site. Similarly, in the case of 1T-FeAs, three different paths have been selected for the migration of Li from one stable Fe2'-site to adjacent Fe2'-sites via the Fe-As bonds, marked path I; path II for Li migration through Fe1'-site and As'-site marked as path III, as shown in **Figure 6(b).**

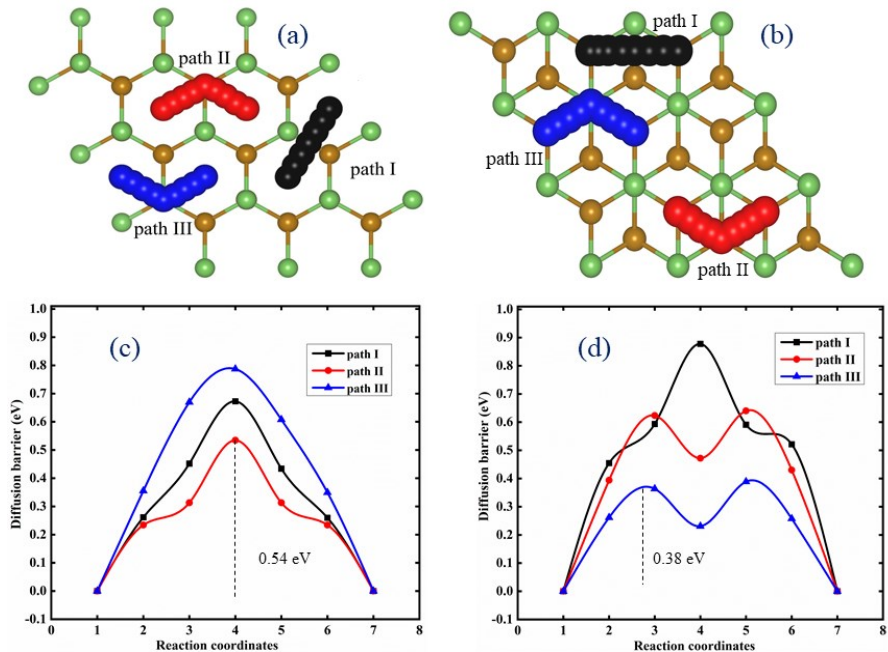

**Figure 6:** *(a) and (b) depict the diffusion path for Li atoms migration (c) and (d) show the diffusion barrier in 1H-FeAs and 1T-FeAs, respectively.*

Our calculations show that Li faces a high diffusion barrier when following path-III in the case of 1H-FeAs due to heightened energy barriers. When travelling along path-II, the 1H-FeAs monolayer exhibits remarkably low diffusion barriers for Li of 0.54 eV. In contrast, Li faces a low barrier of 0.38 eV along path-III and a high barrier of 0.87 eV along path-I for 1T-FeAs. This small diffusion barrier for Li in the 1T-FeAs monolayer facilitates faster Li transport and higher charge/discharge rates compared to common commercial anode materials like graphite[48] and TiO$_2$[49] polymorphs, which typically require overcoming diffusion barriers in the range of 0.35–0.65 eV. Hence, it is important to emphasize that the 1T-FeAs monolayer exhibits outstanding ion mobility compared to other transition metal dichalcogenides (TMDs), where Li diffusion barriers range from 0.5 to 2.40 eV[50-52]. Consequently, the 1T-FeAs monolayer promotes rapid transport of Li atoms, indicating enhanced rate capability for Li-based batteries.

### 3.3.4. Theoretical storage capacity and open-circuit voltage

Determining the average open circuit voltage (OCV) and storage capacity (C) is crucial in understanding the charge-discharge mechanism of rechargeable Li-ion batteries. To illustrate this mechanism for a FeAs monolayer-based Li-ion battery, the following half-cell reaction has been employed:

$$FeAs + xLi^{1+} + xe^- \leftrightarrow Li_xFeAs \quad (4)$$

The OCV has been estimated by calculating the Gibbs free energy ($\Delta G$) change for **eq. (4)** half-cell reaction.

$$\Delta G = \Delta E + P\Delta V - T\Delta S \quad (5)$$

In **eq. (5)** the second term, $P\Delta V$, is considered negligible because there is no remarkable stress on the system during the adsorption of Li on the FeAs surface. Additionally, the entropy term ($T\Delta S$) is approximately 25 meV at ambient temperature, which is very small compared to the average adsorption energy (1-2 eV) of Li atoms. Therefore, the overall change in Gibbs free energy is approximately equal to the net change in internal energy ($\Delta E$), which represents the average adsorption energy ($E_{ave}$). Consequently, the average OCV can be calculated from the average adsorption energy ($E_{ave}$).

$$OCV = -E_{ave}/xne \quad (6a)$$
$$E_{ave} = E_{M_xFeAs} - E_{FeAs} - x\, E_{Li} \quad (6b)$$

Where $E_{Li_xBCN}$, $E_{BCN}$, and $E_{Li}$ are the energies of the Li-adsorbed FeAs, pristine FeAs nanosheet, and Li atom, respectively, and n(=1) is the valency of the Li atom. 'x' represents the number of Li atoms, and the maximum value of x can be obtained by gradually increasing the Li concentration on the FeAs monolayer.

The following equation has been used to compute the relevant adsorption capacities:

$$C = xF/W_{M_xFeAs} \quad (7)$$

where $C$, $x$, $F$ and $W_{Li_xFeAs}$ are storage capacity, concentration of Li atom per unit cell of FeAs monolayer, Faraday's constant and molecular mass per unit cell of Li-adsorbed FeAs monolayer, respectively.

Examining the possibility of multi-layer Li adsorption on the anode FeAs surface is important. In this context, multi-layer Li adsorption significantly enhances ion storage capacity. Consequently, we introduced multiple layers of Li atoms onto each FeAs monolayer to ascertain the maximum Li storage capacity. To illustrate this procedure, we employed the gradual deposition of Li atoms on the FeAs surface using the unit cell, which consists of two Fe and two As atoms ($Fe_2As_2$) in both monolayers. Furthermore, the layers of Li atoms on the FeAs follow the same trend as in single-site adsorption with the increased concentration of Li atoms. For x=1, Li sits at the H-site; for x=2, two Li atoms layers below and above on the FeAs sheet at the H site; and for x=4, four layers of Li atoms, with the last two Li atoms occupying the Fe position in 1H-FeAs, whereas four layers of Li occupies the Fe and As on both sides, bottom and top sides. Therefore, the fully relaxed geometry of FeAs phases is maximum up to 4 layers; there are eight atomic levels on the z-axis, as seen in **Figure 7**.

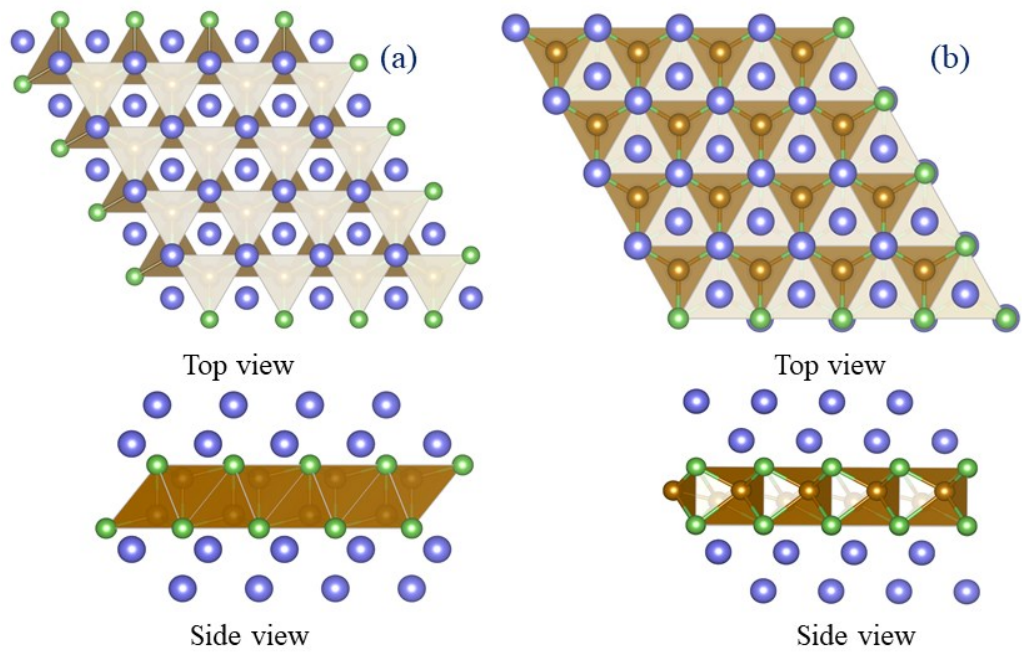

**Figure 7**: *The top and side views of optimization layers of Li atoms on the (a) 1H-FeAs and (b) 1T-FeAs.*

The maximum computed theoretical capacity of Li adsorption for FeAs monolayer is 372 mAhg$^{-1}$ for x=4, which is comparable to commercial using graphite[48] and low dimensional material like MoS$_2$,[53] phosphorene [54] and other TMD material[44, 52, 55]. Both the FeAs phases adsorbed Li atoms up to the four layers. Although the average adsorption energy calculated by increasing the Li concentration will differ for different phases, it is still negative up to x=4, as shown in **Figure 8(c)**. The open-circuit voltage of the electrode material of the battery is an important parameter for power density and affects its performance. A low OCV of the anode is required for a maximal operating voltage of a LiB linked to the cathode. Because the OCV is computed by the change in the chemical potential of the Li atom, it is important to identify intermediate structures during the charging/discharging process to evaluate the OCV of electrode material.

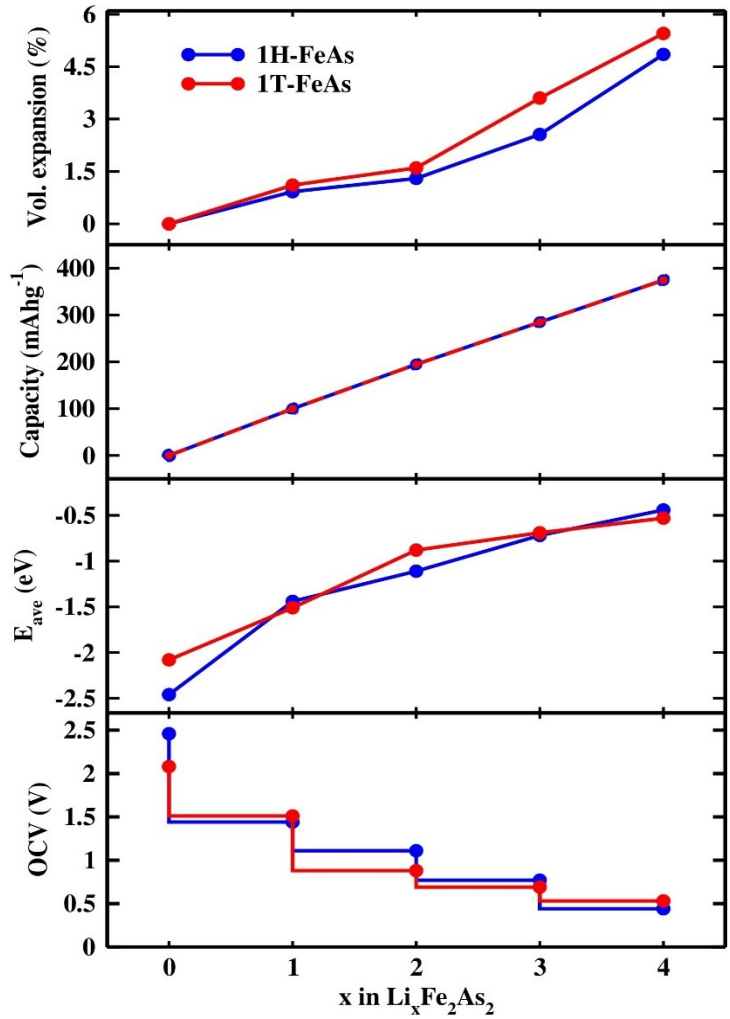

**Figure 8:** *(a) Volume expansion, (b) theoretical capacity, (C) average adsorption energy ($E_{ave}$), (d) open circuit voltage plots with the concentration of the Li ions in 1H-FeAs and 1T-FeAs phases.*

The computed OCV has been plotted in **Figure 8(d)** with the concentration of Li and has the minimum value of 0.61 V for 1T-FeAs and 0.44 V for 1H-FeAs within an acceptable maximum adsorption of Li atoms (x=4) on FeAs phases. These average OCV values for Li lie in the range of commonly used commercial anode materials, specifically 0.11 V for graphite[48] and 1.50-1.80 V for $TiO_2$[49]. In summary, when considering key electrochemical parameters like average open-circuit voltage and specific capacity, both phases of FeAs, 1T-FeAs and 1H-FeAs, show similar performance as potential lithium battery anode materials. However, it is important to note that the 1T-FeAs phase stands out with better ionic conductivity in terms of diffusion barriers.

### 3.3.5. Volume expansion during the lithiation of FeAs monolayer

During the charging and discharging processes, changes in the volume of anode materials can result in fracturing these materials, ultimately leading to a significant loss in capacity. To assess the impact of Li adsorption on the volume change of FeAs monolayers, we examined the in-plane expansion of the FeAs single layer. Li-FeAs comprise eight atomic layers stacked along the z-axis as Li-Li-As-Fe-Fe-As-Li-Li for both phases. The optimized unit-cell of Li adsorbed 1H-FeAs and 1T-FeAs 2D sheets maintain their lattice symmetry of pristine FeAs trigonal *p3m1*. As expected, the lattice parameter increases with the concentrations of Li adsorption. The in-plane expansion of FeAs phases with the Li concentration has been plotted in **Figure 8(a).** The maximum in-plane lattice expansions observed on

1T-FeAs and 1H-FeAs are 5.57% and 4.85% for maximum Li adsorption, respectively. The volume changes observed in the FeAs phases during lithiation/de-lithiation processes are notably less than those in graphite, which can undergo up to a 10% expansion/contraction. This implies that the expansion and contraction of the FeAs monolayer during Li atom intercalation and deintercalation are not a significant concern regarding volume changes.

### 3.4. Effect of Li decoration on the magnetic properties of FeAs phases

We compared the total energies of the possible magnetic state of the 1H-FeAs structure to establish the ground-state magnetic order. It has been found that the energy difference between the Ferromagnetic (FM) state with antiferromagnetic (AFM) and non-magnetic (N) states is 109.5 meV and 134.1 meV, respectively. It is conclusive that the single-layer 1H-FeAs ground state is FM. According to our calculations, each primitive cell has an integer magnetic moment of around 4$\mu_B$. The local magnetic moment per Fe atom is 2$\mu_B$, which signifies that each Fe atom exists in its low spin state. Similarly, the 1T-FeAs phase has an FM ground state compared to other spin-polarized and unpolarized states, as reported in **Table 2**.

**Table 2**: *The energy difference between the magnetic and non-magnetic states for the pristine and Li-adsorbed at the most favourable site of FeAs monolayers are reported.*

| Phases | AFM-N(meV) | FM-N(meV) | AFM-FM(meV) | Ground state |
|---|---|---|---|---|
| FeAs(1H) | -24.53 | -134.09 | 109.55 | FM |
| Li-FeAs(1H) | -141.64 | -60.93 | -80.71 | AFM |
| FeAs(1T) | -196.85 | -437.81 | 240.95 | FM |
| Li-FeAs(1T) | -2.10 | -2.95 | -0.90 | FM |

In the meantime, Li decorated at the top of the H-site of 1H-FeAs phase converts the magnetic state from FM to AFM. Our investigation examined the potential magnetic states of 1H-LiFeAs with spin-polarization and determined that its ground state is AFM. The ground state of 1H-FeAs is FM with a magnetic moment of 2 $\mu_B$ per Fe atom, while Li-FeAs adopts the AFM ground state with a magnetic moment of 1.3 $\mu_B$. In contrast, 1T-FeAs monolayers sustain their magnetic state when Li is decorated at the top of the Fe2' atom. The magnetic moment of 1T-LiFeAs is approximately 2.6 $\mu_B$ per unit cell, with the local magnetic moment per Fe atom being 1.3 $\mu_B$.

This observation indicates that each Fe atom exists in its low spin state, similar to 1H-FeAs. Furthermore, the total energy for the AFM configuration in 1T-LiFeAs is comparable to that of the other magnetic configurations. For instance, the energy difference between AFM and the FM and N states is approximately 0.080 eV and 0.141 eV, respectively. The ground state energy order for the 1H-LiFeAs system is AFM > FM > N. In addition, the magnetic anisotropic energy (MAE) is determined by calculating the difference in total energy between in-plane and out-of-plane spin states in a magnetic system, denoted as $\Delta E = |E_\rightarrow - E_\uparrow|$. For 1H-LiFeAs, this value is predicted to be 3.64 meV. A positive MAE value in 1H-LiFeAs indicates that the easiest axis for magnetization lies out of the plane. Notably, 1T-LiFeAs maintain the stability order observed in pristine 1T-FeAs phases. The primary difference lies in the conversion from high-spin states to low-spin states. Hence, it can be concluded that lithiation of 1H-FeAs changes its magnetic state and induces a transition from high-spin to low-spin states. On the other hand, the 1T-FeAs phase maintains its original ground state, but there is a transition from high-spin to low-spin configurations.

### 4. Conclusions

Using DFT calculations, we thoroughly investigated the structural, electronic and magnetic properties of 1T-FeAs and 1H-FeAs monolayers. The stability of both the monolayers is confirmed by the absence of negative frequency modes in phonon dispersion spectra and negative cohesive energy calculations. The spin polarization electronic band structures show the metallic behaviour for both phases. Further, we evaluated the feasibility of using FeAs-based monolayers as anode materials for Li-ion batteries. A global search has been done to find the most preferable sites for Li adsorption on the surface of FeAs phases. The H-site for 1H-FeAs and Fe2'-site for 1T-FeAs monolayer are energetically favourable sites. The concentration of Li has been incrementally increased based on adsorption energy to determine the maximum specific capacity. The most favourable sites are prioritized, while less favourable ones are considered later in adsorption. Our findings suggest that FeAs monolayers are most suited as an anode material for Li-ion batteries since they can adsorb up to four layers of Li (two above and two below). The greatest specific capacities computed for Li in both FeAs phases are 372 mAhg$^{-1}$. The minimum activation energies for 1H-FeAs atoms are 0.54 eV and 0.38 eV for 1T-FeAs atoms, indicating fast diffusion. Furthermore, the average OCVs of 1H-FeAs are 0.61 V, and 0.44 V for 1T-FeAs monolayer are relatively low. Our calculations indicate that both Fe-based monolayers may be the best anode material for Lithium-ion batteries; however, 1T-FeAs has a lower barrier height than 1H-FeAs, making it an effective choice for use as an anode material in lithium-ion rechargeable batteries.

## Acknowledgement
AK thanks the University Grants Commission (UGC), New Delhi, for financial support in the form of a Senior Research Fellowship (DEC18-512569-ACTIVE). PP thanks DST-SERB for the ECRA project (ECR/2017/003305).

## Data availability
The datasets generated during and/or analyzed during the current study are available from the corresponding author on reasonable request.

## Code availability
Not applicable

## Conflict of interest
On behalf of all authors, the corresponding author states that there is no conflict of interest.